\documentclass{JHEP2}
\pdfoutput=1
\usepackage{amsmath,epsfig}

\title{A non-perturbative study of massive gauge theories}

\preprint{IFIC/13-64}

\author{ M.~Della Morte\footnote{dellamor@ific.uv.es}, P.~Hern\'andez\footnote{m.pilar.hernandez@uv.es}
\\IFIC (CSIC) and Dpto. F\'{\i}sica Te\'orica, \\
Universidad de Valencia, Edificio Institutos Investigaci\'on, \\
Apt.\ 22085, E-46071 Valencia, Spain\\}

\abstract{
We consider a non-perturbative formulation of an $SU(2)$ massive gauge theory on a space-time lattice, which is also 
 a discretised gauged non-linear chiral model.  The lattice model is shown to have an exactly conserved global $SU(2)$ symmetry.
  If a scaling region for the lattice model exists and the lightest degrees of freedom are spin one vector particles with the same
  quantum numbers as the conserved current, we argue that the most general effective theory describing their low-energy dynamics
   must be a massive gauge theory.  We present results
   of a exploratory numerical simulation of the model and find indications for the presence of a scaling region where both a triplet vector and a scalar remain light.}

\begin{document}

\section{Introduction}

The discovery of a Higgs-like particle \cite{Aad:2012tfa,Chatrchyan:2012ufa} has recently  established a new milestone in the experimental quest for
the origin of electroweak symmetry breaking, which amazingly  seems to be well described by the
simplest mechanism thought off in the sixties \cite{Englert:1964et, Guralnik:1964eu,Higgs:1964pj,PhysRevLett.19.1264}. The reason why the Standard Model (SM) is such a good effective theory, if there is new physics, remains however an open question.   Traditional avenues that have been pursued in the past to address the hierarchy problem  such as supersymmetry, extra dimensions or technicolor have  found no support  in experiment. 

Another unsatisfactory aspect of the problem lies in the lack of a  non-perturbative definition of 
 electroweak symmetry breaking and the Higgs mechanism. In particular, it has been known 
for a long time that in a gauge theory defined on a space-time lattice \cite{Wilson:1974sk}, spontaneous symmetry breaking
cannot take place \cite{Elitzur:1975im}. Even in the presence of gauge fixing it has been shown \cite{Frohlich:1981yi, Frohlich:1980gj} that whether or not
a gauge non-invariant 
condensate can get an expectation value depends on how the gauge is fixed~\cite{Caudy:2007sf}.  
This of course does not exclude the possibility 
that a Higgs-like phase does exist where a  continuum limit can be defined and that it resembles the electroweak
sector of the Standard Model, but such connection has not been firmly established. In other words, there is no derivation of
 the  successful perturbative regime of the electroweak sector of the SM from its  non-perturbative definition on a space-time lattice.

In this line, several studies of gauge-scalar theories were performed in the eighties. Some  early references are \cite{Fradkin:1978dv,Lang:1981qg}.  Much effort was devoted to understanding the phase diagram of such lattice theories, in particular trying to search for phase
transitions separating the Higgs and confinement phases. 
It was shown  that   no phase transition line separates these two phases (i.e. the Higgs and confinement phases are continuously connected) 
as a consequence of analiticity \cite{Fradkin:1978dv}.  However, the interesting question of whether a scaling region
could exist in the Higgs phase, and whether the dynamics in this phase is a small perturbation of the trivial scalar model or a non-trivial one has not been settled. An important question is how one could distinguish such a scaling region  from that in the pure gauge theory. It is not expected that the two phases differ qualitatively at distances of the order of the lattice spacing,  but they should definitely differ at long distances, because the propagating physical states are different. Even though confinement might be at work in both cases, in the sense that only colorless states are asymptotic, static charges can be screened in the Higgs phase but not in the confinement phase.

In this paper we reconsider  the simplest of these theories, the gauged non-linear chiral model defined on a lattice,  and argue that its continuum limit within a Higgs phase, if it exists, could be the simplest model of dynamical electroweak symmetry breaking. The two key ingredients are
the standard mechanism of confinement (the  asymptotic states that survive the continuum limit are gauge singlets) and the existence of an exact global symmetry. Assuming the continuum limit or scaling region of this theory exists, the global symmetry constrains 
the structure of the  Wilsonian effective theory to be a massive gauge theory.
We have performed an exploratory lattice simulation of this theory searching for trajectories of constant physics and we find numerical evidence for the presence
of a scaling region in the lattice model with a light massive triplet vector and a light scalar.

The paper is organized as follows. In section \ref{sec:simple} we present a simple argument of why a theory of conserved global currents is a massive gauge theory.
In section \ref{sec:lattice} we present the lattice gauged non-linear chiral model that represents the discretisation of a massive $SU(2)$ gauge theory. We
show the existence of an exact global symmetry and rederive the result of section \ref{sec:simple} from the associated Ward identities. 
In section \ref{sec:numerical}, we present 
the results of a numerical simulation at three different value of the bare coupling and study the scaling of various quantities. 
 Our conclusions and outlook  are presented in \ref{sec:conclu}.

\section{Gauge invariance in an effective theory of conserved currents} 
\label{sec:simple}

 Let us assume that a gauge-scalar model defined on a lattice of spacing $a$ has a scaling region, that is  the lightest excitations with mass $m$ satisfy $m a \ll 1$. Let us furthermore assume that these are spin 1 states that have the same quantum numbers of a conserved current, associated with some exact global symmetry.  Provided these states are lighter than any other states, their dynamics at low momentum must be described by an effective field theory (EFT).

Let us therefore consider an EFT of spin one fields $W^a_\mu$ which transform in the adjoint representation of a global symmetry group that for simplicity we will consider to be $SU(2)$\footnote{In the Standard Model such symmetry would be the custodial symmetry.}.  Let us assume that both Lorentz invariance  and parity are good symmetries. 
We will also make the following fundamental assumption: the fields $W^a_\mu$ are the  conserved currents of such global symmetry. In particular this implies that:
\begin{eqnarray}
\partial_\mu W^a_\mu = 0. 
\end{eqnarray}
We will consider all possible operators compatible with these symmetries up to mass dimension four, assuming that higher dimensional operators would be suppressed by some higher energy scale. 

In principle, the coefficients of such operators are all independent and therefore the theory is not a gauge theory. The most general euclidean Lagrangian can therefore be written in the form:
\begin{eqnarray}
{\mathcal L}_W &=& {1 \over 4} Z_W~  \partial_{[\mu,} W_{\nu]}\cdot \partial_{[\mu,} W_{\nu]} + \alpha~ W_\mu \times W_\nu \cdot \partial_\mu W_\nu + Z_W {m^2_W\over 2}~ W_\mu \cdot W_\mu \nonumber\\ &-& \lambda (W_\mu\cdot W_\mu)^2 - \mu (W_\mu\cdot W_\nu)^2,
\label{eq:lag_global}
\end{eqnarray}
where $Z_W, m_W, \alpha, \lambda, \mu$ are arbitrary.
The global symmetry implies that, under the infinitesimal transformation, $\Omega =e^{i \epsilon(x)}  \in SU(2)$:
\begin{eqnarray}
W_\mu \rightarrow \Omega^\dagger W_\mu \Omega +  i \Omega^\dagger \partial_\mu \Omega,
\end{eqnarray}
the N\"oether current must be conserved:
\begin{eqnarray}
j_\mu^a \equiv {\partial {\mathcal L}_W\over \partial \partial_\mu \epsilon^a(x)} , \;\;\;\; \partial_\mu j^a_\mu = 0.  
\end{eqnarray}
If we
assume the consistency condition that the conserved current associated to the global symmetry is proportional to $W_\mu$:
\begin{eqnarray}
{\partial {\mathcal L}_W\over \partial \partial_\mu \epsilon^a_\mu(x)} \propto W_\mu^a(x),
\label{eq:magic}
\end{eqnarray}
 the following relations must be satisfied
\begin{eqnarray}
\alpha = -4 \lambda = 4 \mu = Z_W.
\label{eq:magicrels}
\end{eqnarray}
while $m_W^2$ is not constrained. After canonically normalizing the kinetic term and defining $g \equiv Z_W^{-1/2}$ we end up with the following Lagrangian:
\begin{eqnarray}
{\mathcal L}_{eff} &=& {1 \over 4}   \partial_{[\mu, } W_{\nu]}\cdot \partial_{[\mu,} W_{\nu]} + g~ W_\mu \times W_\nu \cdot \partial_\mu W_\nu +  {m^2_W \over 2}~ W_\mu \cdot W_\mu \nonumber\\ &+& {g^2\over 4} \left[(W_\mu\cdot W_\mu)^2 - (W_\mu\cdot W_\nu)^2\right] 
\label{eq:lag_gauge}
\end{eqnarray}
 Therefore we find that the most general EFT with the aforementioned properties and including terms with dimension  
 $d\leq 4$ is an $SU(2)$ gauge theory up to  a mass term.

This result can be extended to matter fields straightforwardly. Let us consider for example a  Dirac fermion field, $\Psi$, that transforms in the fundamental of $SU(2)$, the allowed couplings of this fermion up to dimension 4 being:
\begin{eqnarray}
{\mathcal L}_{\Psi} = Z_\Psi ( \bar{\Psi} \gamma_\mu \partial_\mu \Psi + m \bar{\Psi} \Psi) + \delta~ \bar{\Psi} \gamma_\mu W_\mu \Psi  .
\end{eqnarray}
The contribution of this term to the global current is 
\begin{eqnarray}
{\delta {\mathcal L}_\Psi \over \partial_\mu \epsilon^a_\mu(x)} = -i Z_\Psi \bar{\Psi} \gamma_\mu T^a \Psi - \delta  \bar{\Psi} \gamma_\mu T^a \Psi,  
\end{eqnarray}
with $T^a=\frac{1}{2} \sigma^a$, where $\sigma^a$ is a Pauli matrix.
It hence follows that in order to satisfy eq.~(\ref{eq:magic}) we need to require
\begin{eqnarray}
\delta= - i Z_\psi.
\end{eqnarray}
By canonically normalizing the $\Psi$ and the $W_\mu$ fields we therefore obtain a gauge invariant fermion-gauge coupling:
\begin{eqnarray}
{\mathcal L}_{\Psi} =  \bar{\Psi} (\gamma_\mu D_\mu + m) \Psi, 
\end{eqnarray}
with $D_\mu \equiv \partial_\mu - i g W_\mu$. The gauge coupling in the boson self-interactions and the fermion-boson interactions 
are the same. The case of massive fermions with chiral charges will be considered elsewhere. 

We note that, by using a different approach, the same EFT derived here from symmetry arguments, 
has been previously shown in~\cite{Gegelia:2012yq} to provide
the most general Lagrangian describing at low energy the interactions of massive vector bosons coupled to fermions.
  
\section{Massive gauge theories as confining gauged non-linear $\sigma$ models} 
\label{sec:lattice}

We consider  the $SU(2)$-gauged non-linear chiral model, which contains the degrees of freedom corresponding
to the massive gauge bosons and no fundamental scalars. 
It is well-known that such a model defined on the lattice is exactly equivalent to a lattice-regularized $SU(2)$ theory with an explicit mass term. Let us recall how this works. The simplest  lattice action of a massive $SU(N)$ Yang-Mills theory in a cubic lattice of lattice spacing $a$ is given by
\begin{eqnarray}
S_m[U] = -{\beta \over 2N} \sum_{x} \sum_{P}  {\rm tr} \left[ P(x)+ h.c.\right] - {\kappa \over 2} \sum_{x } \sum_\mu { \rm tr} \left[ U_\mu + U_\mu^\dagger\right]\; ,  
\label{eq:sm}
\end{eqnarray}
where $P(x)$ is the elementary plaquette, $U_\mu(x)$ is the link variable,  $\beta \equiv 2 N/g^2_0$, and $\kappa \equiv {\beta\over N} (m a)^2$.  

 In this formulation, the theory looks like it is not gauge invariant, but it can be rewritten as a gauge invariant theory after performing an integration over the gauge orbit. Using the compactness of the group and the invariance of the measure we can write the partition functional as
\begin{eqnarray}
Z &=& \int \prod_x d \Omega(x) ~\prod_{x,\mu} d U_\mu(x)   \exp\left[- S_m[U] \right] = \int \prod_x d\Omega(x) \prod_{x,\mu} d U^\Omega_\mu(x) \exp\left[- S_m[U^\Omega] \right] \nonumber\\
&=&   \int \prod_x d\Omega(x) \prod_{x,\mu} d U_\mu(x) \exp\left[- S_m[U^\Omega] \right] = \int \prod_x d\Omega(x) \prod_{x,\mu} d U_\mu(x) \exp\left[- S[U, \Omega] \right]. \nonumber\\
\end{eqnarray}
where $U_\mu^\Omega$ is the gauge-transformed link variable,
\begin{eqnarray}
U^\Omega _\mu(x)= \Omega^\dagger(x) U_\mu(x) \Omega(x+a \hat{\mu}),
\end{eqnarray}
and 
\begin{eqnarray}
S[U, \Omega] &\equiv& -{\beta\over 2 N} \sum_{x } {\rm tr} \left[ P(x) +h.c.\right] \nonumber \\
&-& {\kappa \over 2} \sum_{x } \sum_\mu {\rm tr} \left[ \Omega^\dagger(x) U_\mu(x) \Omega(x+a\hat{\mu}) + h.c.\right] .
\end{eqnarray}
This action is now invariant under the following gauge transformation
\begin{eqnarray}
U_\mu(x) &\rightarrow & \Lambda(x) U_\mu(x) \Lambda^\dagger(x+a \hat{\mu}),\\
\Omega(x) &\rightarrow & \Lambda(x) \Omega(x).
\end{eqnarray}
In this formulation, the theory contains gauge degrees of freedom coupled to complex scalars, $\Omega \in SU(N)$, on which the gauge transformation acts on the left. It is a discretized version of the gauged non-linear chiral model. For $N=2$ this is also a non-perturbative formulation of an $SU(2)$+ $\lambda \phi^4$ theory, in the limit of an infinite Higgs mass $\lambda \rightarrow \infty$.

We make also the important observation that this lattice theory satisfies site-reflection positivity and therefore unitarity \cite{Osterwalder:1973dx,Osterwalder:1974tc}. 

The theory obviously  reduces to the pure gauge theory in the limit $\kappa\rightarrow 0$ (confined phase), while the scalar degrees of freedom decouple from the gauge 
interactions  in the  na\"ive $\beta\rightarrow \infty$ limit, simplifying to the ungauged  non-linear sigma model. The latter is in the same universality class as the $\lambda \phi^4$ theory and has  a second order phase transition at $\kappa = \kappa_c$, between an ordered (Higgs) phase and a disordered one. For an exhaustive study of this ungauged limit of the lattice model see \cite{Luscher:1988uq} and references therein. 

The phase diagram of this theory as function of $(\beta,\kappa)$ has been studied before \cite{Fradkin:1978dv,Lang:1981qg,Osterwalder:1977pc,Langguth:1985dr}.  For more recent studies see \cite{Campos:1997dc,Caudy:2007sf,Bonati:2009pf,Ferrari:2011aa,Bettinelli:2012qk}. Using the small $\kappa$ limit (high T expansion) at finite $\beta$, it is possible 
to show that the theory falls in the same universality class as the pure $SU(2)$ gauge theory \cite{Forster:1980dg}. There must exist therefore a confinement
phase where the continuum limit is the same as in the pure $SU(2)$ gauge theory.  In this phase, the $\Omega$ fields decouple as heavy degrees of freedom as we approach the continuum limit at $\beta \rightarrow \infty$.  It is reasonable to assume that for sufficiently large $\kappa$, the $\Omega$ fields 
can have long range correlations and remain in the spectrum. That would be a Higgs-like phase.  

For large $\kappa$ and $\beta$ the standard perturbative expansion indicates that the theory is non-renormalizable, but it has some validity as a momentum expansion (in the same sense that
chiral perturbation theory does). To be more precise, considering external momenta such that $m^2 \equiv {\kappa g_0^2\over 2 a^2} \sim p^2 \ll  {\kappa\over a^2}$, the perturbative
expansion is expected to be renormalizable order by order  in  ${p^2 a^2\over \kappa}$.  In the appendix, we present the one-loop corrections to $\beta$ and $\kappa$ in this regime  using the background field method on the lattice. As expected, $m^2 \propto {\kappa \over a^2}$ being a dimensionful coupling 
gets cutoff-scale corrections and needs to be non-perturbatively fine-tuned to remain light in units of the cutoff. On the other hand, within the validity of the momentum expansion, the theory is found to be asymptotically free in $g_0$:
\begin{eqnarray}
\beta(g_0) = -{1\over 16 \pi^2}\left({22 \over 3} - {1\over 12}\right) g_0^3+..., 
\label{eq:beta}
\end{eqnarray}
which corresponds to a scalar contribution to the $\beta$ function which is half of  that of one complex doublet. The same result has been obtained in a different regularisation in \cite{Fabbrichesi:2010xy}.
These one-loop results suggest that  a  scaling region might exist at $\beta = \infty$ for some finite and non-perturbatively fine-tuned value of $\kappa$. Whether that region exists cannot be established within perturbation theory however, since an infinite number of tunings seem necessary to find scaling to arbitrary order  in this expansion.  On the other hand if such a scaling region is found non-perturbatively, the model might provide the simplest model of dynamical symmetry breaking: all 
the low-energy parameters describing the low-energy dynamics would be determined in terms of two or less (if there is asymptotic freedom) bare couplings.  Note that 
this is precisely what happens in the ungauged non-linear model, a tuning of $\kappa$ is all what is required to reach the (trivial) continuum limit, in spite of the fact
that the perturbative (large $\kappa$) expansion indicates otherwise. 

As explained in the introduction, the existence of a scaling region within a Higgs phase  implies that the static potential shows a mixed behavior rising linearly up to some physical distance $r_s$, related to the scale of string breaking, and flattening thereafter. If such a physical scale exists it must satisfy scaling $r_s/a \rightarrow \infty$ as $a\rightarrow 0$.  Another observable that would be distinct in a Higgs phase  is the correlation function of states constructed out
of $\Omega$ fields, which should not have long range correlations in the confinement phase, i.e. $\xi_{\Omega}/a$ remains finite as $a\rightarrow 0$. 

An essential property of   this lattice model is that it has an exact global symmetry under the transformation
\begin{eqnarray}
\Omega(x) &\rightarrow & \Omega(x) \Lambda, \;\; \Lambda \in SU(N). 
\label{eq:glosym}
\end{eqnarray}
There is an {\it exactly} conserved vector current associated to this global symmetry which is given by
\begin{eqnarray}
V^a_\mu \equiv i {\kappa \over 2}  {\rm Tr}[ \Omega(x)^\dagger U_\mu(x) \Omega(x+\hat{\mu}) T^a] + h.c. 
\label{eq:cvec}
\end{eqnarray}
This current satisfies
\begin{eqnarray}
\hat{\partial}_\mu V^a_\mu = 0, 
\end{eqnarray}
where $\hat{\partial}_\mu \Omega(x) \equiv \Omega(x+\hat{\mu}) - \Omega(x)$ is the forward lattice derivative. 

  Let us assume that a scaling region exists and that  only colorless
states are asymptotic. One such state carries  the  quantum numbers of the 
conserved current, $V^a_\mu$.
The EFT for such state in the continuum limit must look like a massive gauge theory, as described in the previous section, since it satisfies all the properties required. In particular,
the interactions of this state (with itself and other non-singlet fields) are controlled by the  effective gauge coupling constant, $g$, which need not be related in any simple way to the bare coupling $g_0$. 

The relations of eq.~(\ref{eq:magicrels}) that underlie the effective gauge symmetry can in fact be derived also from the Ward identities (WI) associated to the exact global symmetry as we show next. Such WI also provide a non-perturbative definition of the effective coupling $g$. 

We note the connection of this formulation with the so called hidden local symmetry construction of non-linear sigma models \cite{D'Adda:1978uc,D'Adda:1978kp,Balachandran:1978pk} that was applied to describe light vector mesons in chiral perturbation theory \cite{Bando:1984ej,Georgi:1989gp} and also to models of electroweak symmetry breaking \cite{Casalbuoni:1986vq}. Here the hidden local symmetry 
is a true local symmetry, the only fundamental gauge symmetry.

 \subsection{Ward Identities}

The existence of an exactly conserved global symmetry implies that the following relation must hold for any operator $O$ 
\begin{eqnarray}
\langle -\delta S ~O \rangle + \langle \delta O \rangle = 0, 
\end{eqnarray}
where $\delta S, \delta O$ are the variation of the lattice action and operator respectively under an infinitesimal and local 
symmetry transformation, $\Lambda(x)= e^{i T^a \epsilon^a(x)}$. The variation of the action is 
\begin{eqnarray}
\delta_\epsilon S[U,\Omega] = - \sum_{x,\mu,a} V_\mu^a(x) \hat{\partial}_\mu \epsilon^a(x),
\end{eqnarray}
where $V_\mu^a(x)$ is the conserved vector current of eq.~(\ref{eq:cvec}).

Let us first consider the operator $O(y,z) \equiv V_\mu^a(y) V_\nu^b(z)$ and the following infinitesimal transformation (this is the method of \cite{luscher:1998pe} and references therein):
\begin{eqnarray}
\epsilon^a(x) = \left\{\begin{array}{ll}
\epsilon^a, & x \in R\\
0, & x \notin R
\end{array}\right.
\end{eqnarray}
where $R$ is the region $0 \leq t \leq T$. We assume that $y \in R$ (i.e. $0 < y_0 < T$) while $z \notin R$, for example $z_0 > T$. The boundaries of $R$, that we denote 
by $\partial R$,  are therefore
infinite hyperplanes at constant $t=0$ and $T$. The transformation of the field is easily derived from the transformation of the current
\begin{eqnarray}
\delta_\epsilon V^a_\mu(x) =  \epsilon^{abc} \left[\epsilon^b(x) ~V^c_\mu(x) +{1\over 2} \hat{\partial}_\mu \epsilon^b(x) ~ÊV^c_\mu(x)\right]-{1 \over 4} V_\mu^0(x) \hat{\partial}_\mu \epsilon^a(x),
\end{eqnarray}
where $V^0_\mu$ is a singlet under the global symmetry:
\begin{eqnarray}
V^0_\mu(x) \equiv  {\kappa \over 2} {\rm Tr}[ \Omega(x)^\dagger U_\mu(x) \Omega(x+\hat{\mu})  + h.c.].
\label{eq:h}
\end{eqnarray}

The lattice  Ward identity in this case results in the following relation
\begin{eqnarray}
-\epsilon_{abc}  \sum_{\mathbf{x}}  \langle  \left( V_0^c(T,\mathbf{x}) -V_0^c(0,\mathbf{x}) \right)V_\mu^a(y) V^b_\nu(z) \rangle  = 2 \langle V_\mu^d(y) V_\nu^d(z) \rangle,  
\end{eqnarray}
where we have used that $\partial_\mu\epsilon(y) = 0,  y \notin \partial R$. Note that all the operator insertions $y, z$ and $(0,\mathbf{x})$/$(T,\mathbf{x})$ are far apart, and therefore both sides of the equation should match the continuum correlation functions up to a field renormalisation. In the continuum theory, we denote the canonically-normalised field that represents the vector particle 
by $W_\mu^a$.   The Ward identity in terms of continuum fields is therefore 
\begin{eqnarray}
-\epsilon_{abc}  \int_{\mathbf{x}}  \langle \left( W_0^c(T,\mathbf{x}) -W_0^c(0,\mathbf{x}) \right)W_\mu^a(y) W^b_\nu(z) \rangle  = 2 Z_V^{-1/2} \langle W_\mu^d(y) W_\nu^d(z) \rangle,  
\label{eq:wiw}
\end{eqnarray}
where $Z_V$ can be obtained from the large distance behavior of the Euclidean two-point function
\begin{eqnarray}
\int d^4 x e^{i q x} \langle V^a_\mu(x) V^b_\nu(0)\rangle = {Z_V \delta_{ab} (g_{\mu\nu} + q_\mu q_\nu/m^2_W) \over q^2 +m^2_W } \equiv Z_V \delta_{ab} \Delta_{\mu\nu}(q)\;. 
\label{eq:vv}
\end{eqnarray}
We use the normal convention and define $Z_V^{1/2} \equiv m_W F_W$. 

The EFT that represents the continuum limit is the most general renormalizable theory that satisfies the exact global symmetry and therefore is of the general form in eq.~(\ref{eq:lag_global}) after a canonical normalisation of the field. The three-W coupling,  $g\equiv \alpha Z_W^{-3/2}$, and the two four-W couplings $\lambda Z_W^{-2}$ and $\mu Z_W^{-2}$ are in principle unrelated.   
Let us assume that these couplings are small in the EFT and let's evaluate the two sides of this equation at leading order in perturbation theory. 
 
By restricting our attention to the case $\mu=\nu=0$ and evaluating the three-point function at tree level, we obtain
\begin{eqnarray}
&&\epsilon^{abc} \int_{\mathbf x}\langle  \left( W_0^c(T,\mathbf{x}) -W_0^c(0,\mathbf{x}) \right)  W_{0 }^a(y) W^b_{0 }(z) \rangle \nonumber\\
&=&   {3 g \over m_W^2}  \int_{\mathbf{q}} {{\mathbf q}^2/m_W^2  \over    \sqrt{ {\mathbf q}^2+ m_W^2 } }~e^{i {\mathbf q} ({\mathbf z} -{\mathbf y})}~e^{-\sqrt{{\mathbf q}^2+ m_W^2} (z_0 - y_0)}
\nonumber\\
&=& - { 2g \over m_W^2} \langle     W_{0 }^d(y)  W^d_{0 }(z) \rangle,
\end{eqnarray}
and therefore from eq.~(\ref{eq:wiw}) we get
\begin{eqnarray}
{1 \over \sqrt{Z_V}} = {g\over m^2_W} \rightarrow g = {m_W \over F_W}, 
\end{eqnarray}
which provides a definition of the effective coupling at this order. Note that a weak coupling requires $g \sim {\mathcal O}(1)$, which seems to be the natural 
value for a ratio of dynamically generated scales such as $m_W$ and $F_W$, even if the underlying theory is strongly coupled.

Let us now consider the Ward identity where $O \rightarrow V^b_\mu(y) V^c_\nu(z) V^d_\sigma(u)$ and  a local global transformation at $x$. We assume that all points $x,y,z,u$ are far apart of each other. 

The Ward identity  reads
\begin{eqnarray}
\langle \hat{\partial}_\rho V^a_\rho(x) V^b_\mu(y) V^c_\nu(z) V^d_\sigma(u)\rangle = 0. 
\end{eqnarray}
This correlation function should also match its continuum counterpart, once we substitute  $V$ by $W$. 

Let us first consider  the case $a=b=c=d$ fixed and let us compute this at leading order in perturbation theory. 
Due to the color structure there is no contribution from the three-point coupling $g$ at tree level. The result is  
\begin{eqnarray}
&&\langle \hat{\partial}_\rho W^a_\rho(x) W^a_\mu(y) W^a_\nu(z) W^a_\sigma(u)\rangle \nonumber \\
&=&  - 8  (\lambda+\mu) Z_W^{-2} \int_{k,p,q} (k+q+p)_\tau ( \Delta_{\mu\tau}(p) \Delta_{\nu\alpha}(q) \Delta_{\sigma\alpha}(k) \\
&+&\Delta_{\mu\alpha}(p) \Delta_{\nu\tau}(q) \Delta_{\sigma\alpha}(k)+ \Delta_{\mu\alpha}(p) \Delta_{\nu\alpha}(q) \Delta_{\sigma\tau}(k)) e^{i p (y-x)} e^{i q (z-x)} e^{i k (u-x)}\;. \nonumber
\label{eq:aaaa}
\end{eqnarray}
For finite separations between all the points $x,y,z,u$, the integral is well defined and does not vanish, therefore the coefficient must vanish, i.e.
\begin{eqnarray}
\lambda= -\mu. 
\label{eq:fpc}
\end{eqnarray}

Finally we  consider  the case $a=b$, $c=d$ and $a\neq c$. In this case both the three-point coupling $g$ as well as $\lambda$ and $\mu$ contribute at tree level. By imposing the condition eq.~(\ref{eq:fpc}) we get 
\begin{eqnarray}
&&\langle \hat{\partial}_\rho W^a_\rho(x) W^a_\mu(y) W^b_\nu(z) W^b_\sigma(u)\rangle \nonumber \\
&=&  - (4 \lambda Z_W^{-2}+ g^2) 
  \int_{k,p,q} (k+q+p)_\tau ( 2 \Delta_{\mu\tau}(p) \Delta_{\nu\alpha}(q) \Delta_{\sigma\alpha}(k) \\
&-&\Delta_{\mu\alpha}(p) \Delta_{\nu\tau}(q) \Delta_{\sigma\alpha}(k)- \Delta_{\mu\alpha}(p) \Delta_{\nu\alpha}(q) \Delta_{\sigma\tau}(k)) e^{i p (y-x)} e^{i q (z-x)} e^{i k (u-x)} \;. \nonumber
\label{wiwww}
\end{eqnarray}
Also in this case the integral gives a non-vanishing contribution and therefore the coefficient must vanish
\begin{eqnarray}
\lambda Z_W^{-2} = -{g^2 \over 4}. 
\end{eqnarray}
We hence recover  the relations of eq.~(\ref{eq:magicrels}) and the gauge-invariant effective Lagrangian, eq.~(\ref{eq:lag_gauge}).

 The relations eq.~(\ref{eq:magicrels}) are derived from a tree level evaluation of the Ward identities, while we know that in the theory with only the massive 
$W_\mu^a$ fields unitarity breaks down at high energies.  The Born approximation should represent correctly the low-energy dynamics of the theory, provided
$g$ is small and there are no other light degrees of freedom closeby.

This means that we have implicitly assumed that the only light degrees of freedom below the cutoff are the $W_\mu^a$. However, it is likely that other light states exist that might contribute to the Ward identity amplitudes, for example, a light scalar $H$, that might unitarize the theory if the couplings to the $W$ are those of the Standard Model.  An interpolating field that could couple to such a state is that in eq.~(\ref{eq:h}), which is a singlet under the global symmetry. Accordingly the only allowed $d \leq 4$ couplings of this state  could be
\begin{eqnarray}
{\mathcal  L}_{H} = {1 \over 2} \partial_\mu H \partial_\mu H - V(H) - \lambda_{HWW} H W_\mu \cdot W_\mu - \lambda_{HHWW} H^2 W_\mu \cdot W_\mu\;,
\end{eqnarray}
where $V(H)$ is the standard quartic potential. 
If the $H(x)$ field is invariant under the local symmetry, the $HW$  couplings would be forbidden by eq.~(\ref{eq:magic}). In fact such couplings would give 
at LO uncanceled contribution to the right-hand side of eq.~(\ref{wiwww}). However such contribution can be reabsorbed in a redefinition of the conserved
current in the EFT. The matching of the conserved current in the presence of the $\lambda_{HWW}$ is modified to:
\begin{eqnarray}
V_\mu^a \rightarrow W_\mu^a + 2 {\lambda_{HWW}\over m_W^2} H W_\mu^a, 
\end{eqnarray}
which is conserved up to higher order corrections.
The global symmetry does not seem to constrain the scalar to vectors couplings, although we know that they are fixed by the requirement of perturbative unitarity if no other 
light state exists. 

\section{Numerical Results}
\label{sec:numerical}

We have performed an exploratory study of the scaling properties of the lattice model defined in eq.~(\ref{eq:sm}). This is essentially a standard gauge theory with the
 Wilson action and the $\kappa$ term, which
is the sum of the trace of the link variables. The heatbath and over-relaxation algorithms can be used straightforwardly by 
adding to the staples a term proportional to the identity. Periodic boundary conditions are imposed in time and space.

We have measured the vector and scalar zero-momentum two-point functions, that is the correlators:
\begin{eqnarray}
V(t) &=& -\kappa^2 \sum_{\vec{x},k,a} \langle {\rm Tr}[U_k(x) T^a] {\rm Tr}[U_k(0) T^a] \rangle\;, \\
S(t) &=& \kappa^2 \sum_{\vec{x}} \langle \sum_\mu {\rm Tr}[U_\mu(x)] \sum_\nu {\rm Tr}[U_\nu(0)] \rangle\;,
\end{eqnarray}
where $\mu, \nu=0,1,2,3$ and $k=1,2,3$. 
For the scalar, we used  APE smeared links~\cite{Albanese:1987ds}, whereas in the case of the vector we considered both, smeared sources to extract the mass, and unsmeared ones to get
$F_W$, since the exactly conserved current is the unsmeared one. From these two correlators we can extract the physical quantities $m_H$, $m_W$ and $F_W$ using standard definitions (see e.g.~\cite{Langguth:1985dr} for the effective mass).

\begin{figure}
\begin{center}
\includegraphics[width=10cm, angle=0]{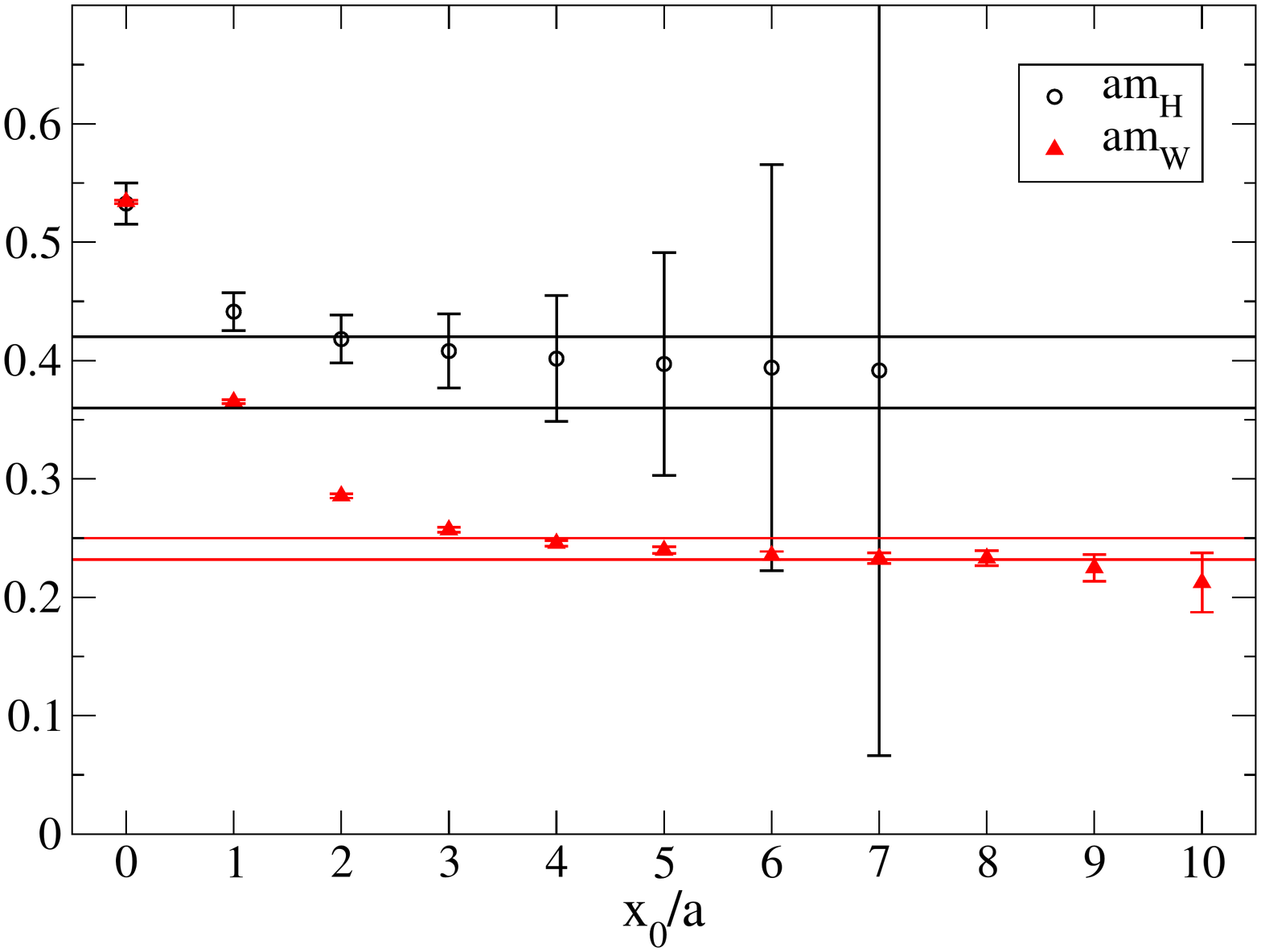}
\caption{Effective mass plateaux from the scalar and vector APE smeared correlators at $\beta=2.55$ and the reference point in $\kappa$.
Bands indicate our final estimates. Errors  account for autocorrelation effects through the method described in~\cite{Wolff:2003sm}.}
\label{fig:effmass}
\end{center}
\end{figure}

The goal is to consider values of the bare parameters that keep some physical quantity fixed. We have considered three values of $\beta$ and $L/a$
starting at $\beta =2.3$, $L/a=16$  and aiming at a change in the lattice spacing by a factor 2-3, guided by the perturbative formula in eq.~(\ref{eq:beta}). The volume
is roughly fixed in physical units. 
In each case we have tuned $\kappa$ to some reference value, $\kappa_{ref}(\beta)$. Inspired by the discovery of the Higgs-like particle, we have chosen 
to keep the ratio $m_H/m_W$ fixed to $\approx 1.5$ (within $\sim 10$\% errors). The results at $\kappa_{ref}(\beta)$  are summarised
in Table~\ref{tab:results}. Some illustrative effective mass plot is shown in Figure~\ref{fig:effmass}. 
Plateaux can be clearly identified, although the noise to signal ratio grows exponentially in time as expected.
In the future we may adopt the algorithm in~\cite{DellaMorte:2008jd,DellaMorte:2010yp} to overcome the problem.
Within our errors we do not see a sign of other nearby bound states and a two state fit of the effective masses
indicates that the first excited state is at the cutoff scale. Of course
it would be interesting to search for such states more thoroughly. 

\begin{table}
\begin{center}
\begin{tabular}{ccccccc}
\hline\\[-2.5ex]
$\beta$ & $\kappa_{ref}$ & $L^3 \times T$ & $am_{H}$ & $am_{W}$& $aF_{W}$ & $N_{\rm meas}/10^6$ \\
\hline
$2.3$   & $0.405$ & $16^3 \times 16$ & $ 0.65(2)$  & $0.455(5)$ & $0.146(2)$ & 5.4 \\
 $2.55$  &  $0.368$ &  $24^3 \times 24$ & $  0.39(3)$  &  $0.241(9)$  &  $0.081(2)$ &  2.8  \\
 $2.75$  & $0.356$ & $36^3 \times 36$ & $  0.30(4)$ &$ 0.174(6) $ & $ 0.060(2)$ & 2.0  \\
\hline
\end{tabular}
\caption{Lattice parameters and estimates for $a m_H$, $a m_W$ and $a F_W$.}
\label{tab:results}
\end{center}
\end{table}

\begin{figure}
\begin{center}
\includegraphics[width=10cm,angle=0]{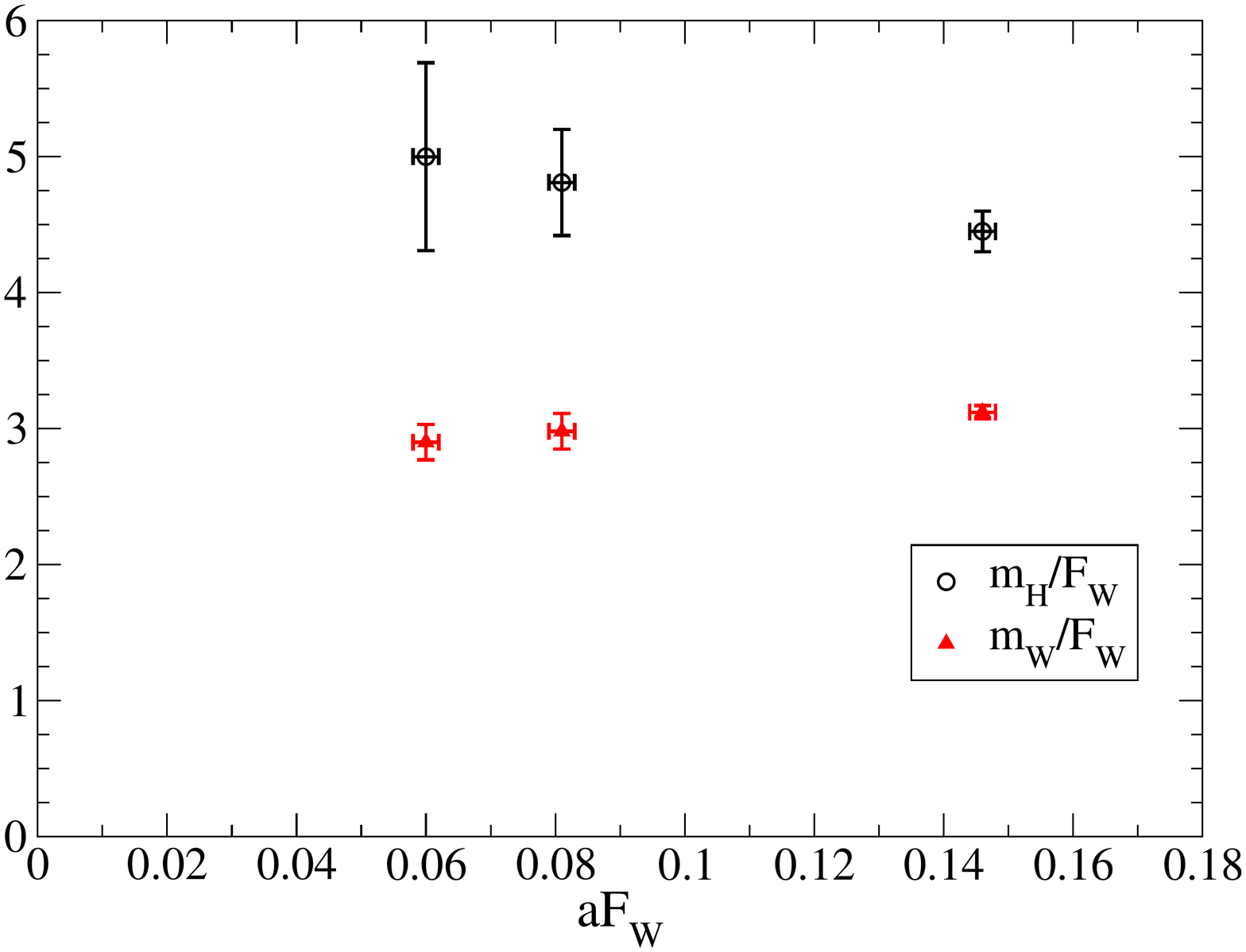}
\caption{Ratios $m_H/F_W$ and $m_W/F_W$ as a function of $a$ in units of $F_W$.}
\label{fig:ratio}
\end{center}
\end{figure}

\begin{figure}
\begin{center}
\includegraphics[width=10cm,angle=0]{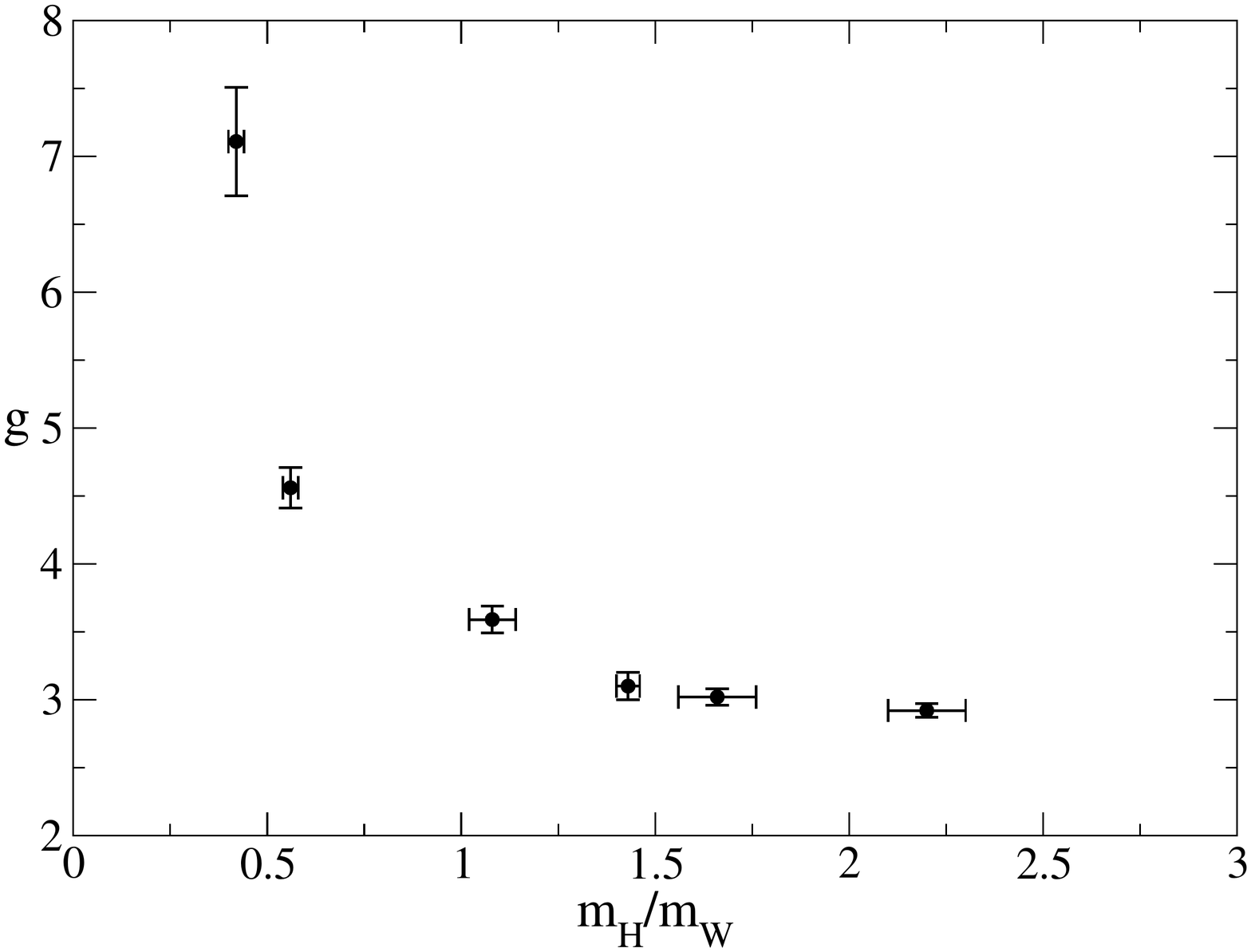}
\caption{Results for the coupling $g$ vs $m_H/m_W$ at $\beta=2.3$, $L/a=16$.
Data from left to right correspond to $\kappa=0.395,\; 0.397,\; 0.400,\; 0.405,\; 0.409$ and $0.413$ and are based on ${\mathcal O}(10^6)$
 measurements at each value of $\kappa$.}
\label{fig:gvsmrat}
\end{center}
\end{figure}

\begin{figure}
\begin{center}
\includegraphics[width=10cm,angle=0]{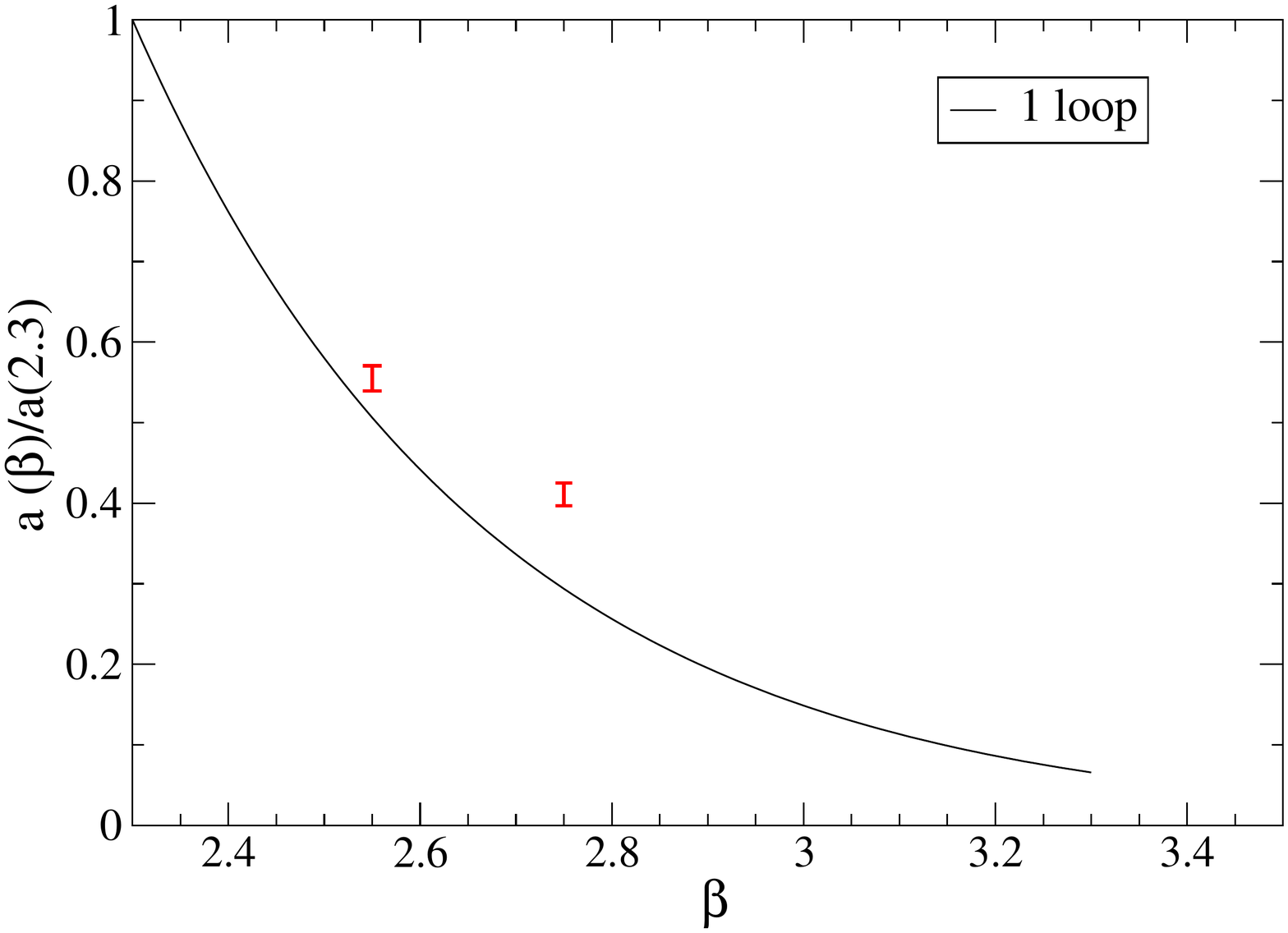}
\caption{Ratio of $[a(\beta) F_W]/[a(2.3) F_W]$ as a function of $\beta$ compared with the one-loop expectation.}
\label{fig:a}
\end{center}
\end{figure}

There seems to be indication of  scaling at this $\kappa_{ref}(\beta)$ as $\beta$ grows: increasing $\beta$ the correlation lengths get larger in lattice units ($a m_W$ drops by a factor $\sim 2.7$  between the coarser and the finer lattices). Furthermore 
the dimensionless ratios $m_W/F_W$, $m_H/F_W$ (only one of them is independent)  remain roughly constant as shown in Figure~\ref{fig:ratio}. 
The ratio $m_W/F_W$ gives the effective coupling, $g$. We can see that the value is however very large  (compared to the $g\sim 0.7$ in the Standard Model). The tuning of $\kappa$ could have also been done in such a way as 
to fix this ratio instead, which might actually be more precise numerically. It would be interesting to see how small a value for this ratio can be achieved
within the scaling region and how it correlates with the ratio of scalar to vector mass. 
A preliminary study at $\beta=2.3$, $L/a=16$ is summarized in Figure~\ref{fig:gvsmrat}. 
The coupling $g$ appears to vary quite rapidly as a function of $m_H/m_W$, especially for small values of the mass ratio, but seems to reach a plateau for large values.
A more extensive study will be performed elsewhere.
The change in the lattice spacing, as extracted from the ratio of $a F_W$, 
is shown in Figure~\ref{fig:a} where it is compared with the one-loop expectation from eq.~(\ref{eq:beta}). It seems that the running is slower than predicted by
 one-loop perturbation theory at these values of $\beta$.

We have also measured correlators of Polyakov loops with APE smearing, from which we have extracted the static potential in the standard way
(which assumes the overlap of the Polyakov loop with the ground state potential to be 1):
\begin{eqnarray}
V(r) = -{1 \over T} \log C_{PP}(r)_{\rm connected}.
\end{eqnarray}

\begin{figure}
\begin{center}
\includegraphics[width=7.cm,angle=0]{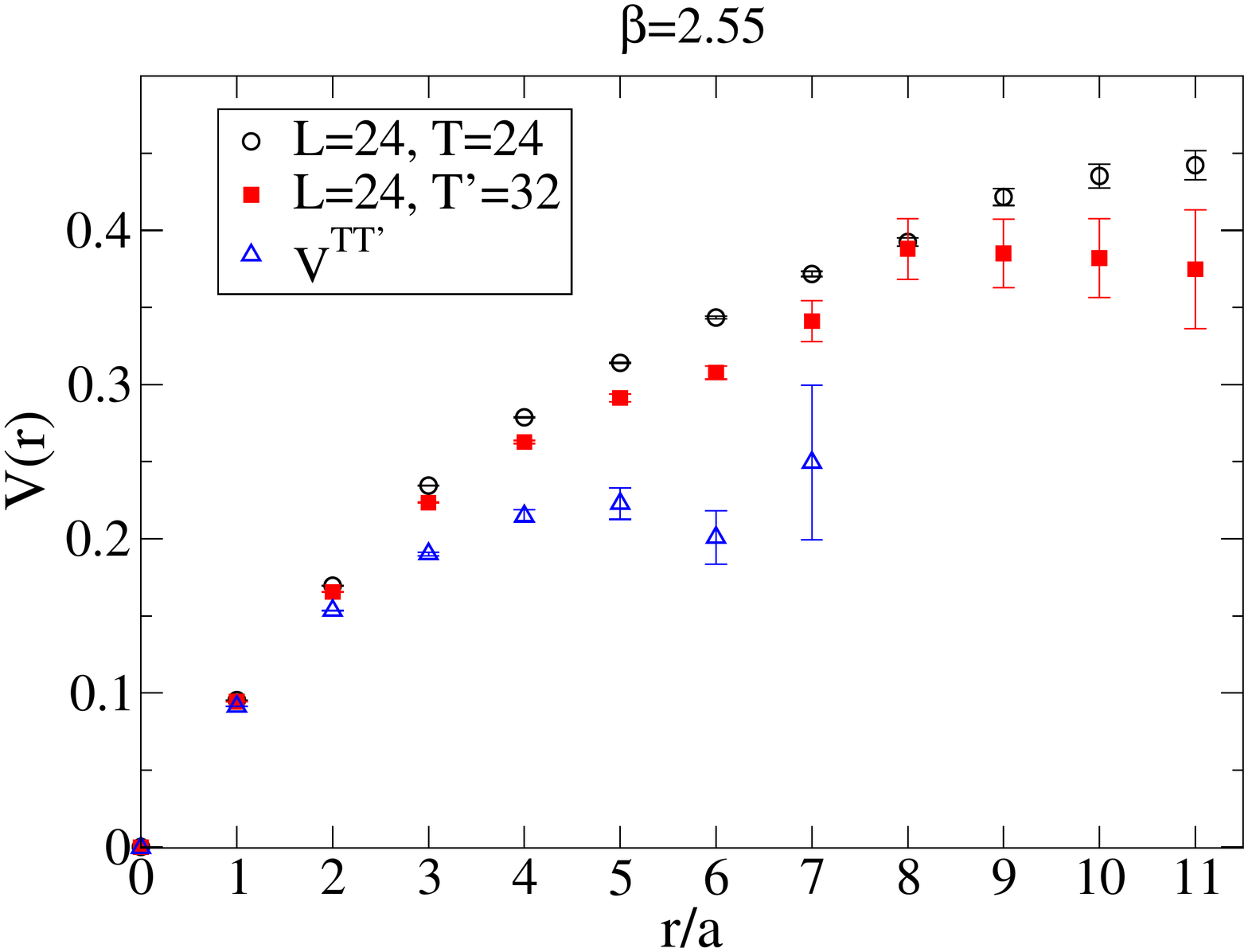} \includegraphics[width=7cm,angle=0]{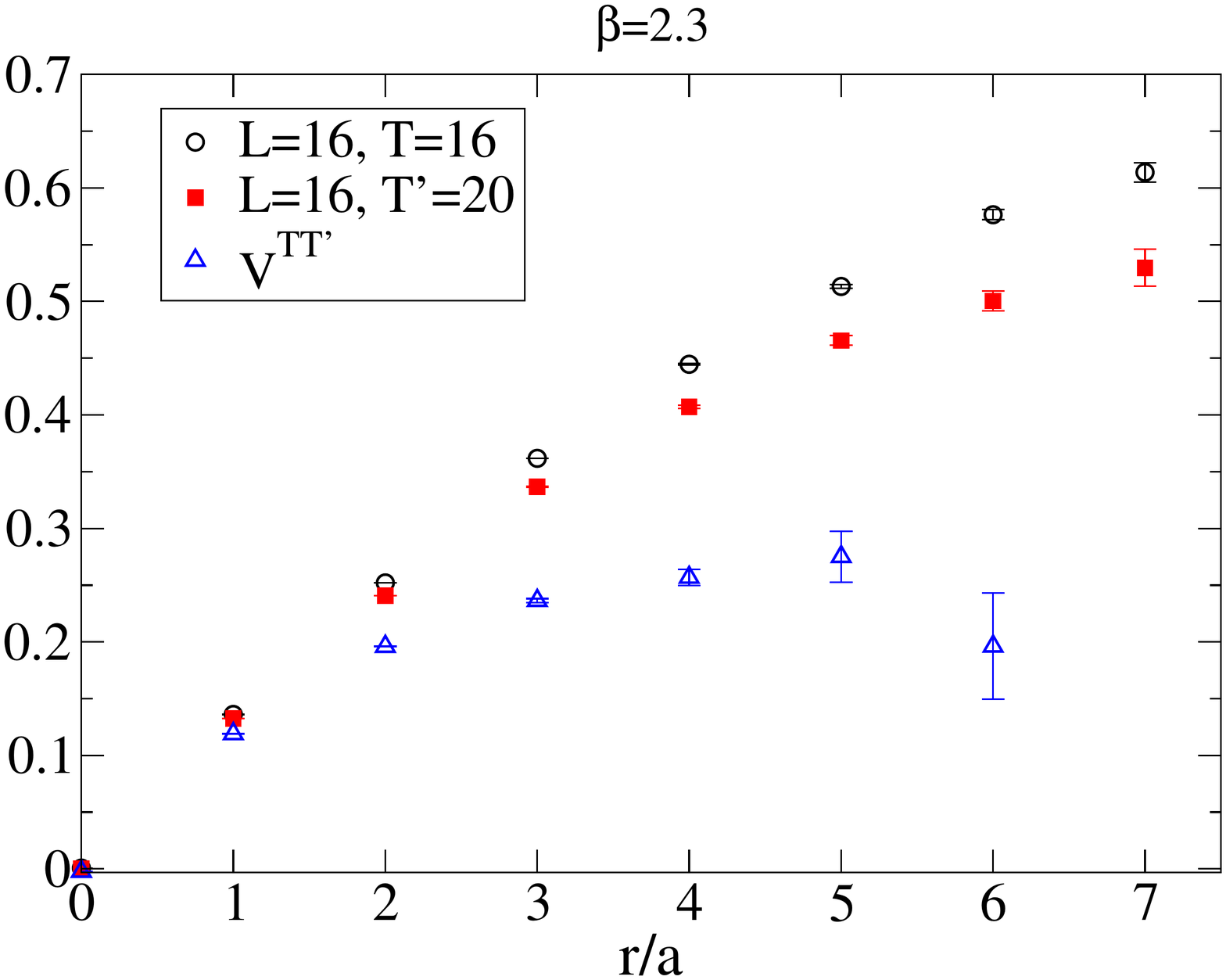}
\caption{Left: static potential extracted from the Polyakov correlator at $\beta=2.55$ for two time extents 
$T=24$, $T'=32$ together with the potential extracted from the ratio (indicated as $V^{\rm TT'}$).
Right: same at $\beta=2.3$ for 
$T=16$, $T'=20$.}
\label{fig:potential}
\end{center}
\end{figure}

\begin{figure}
\begin{center}
\includegraphics[, width=10cm,angle=0]{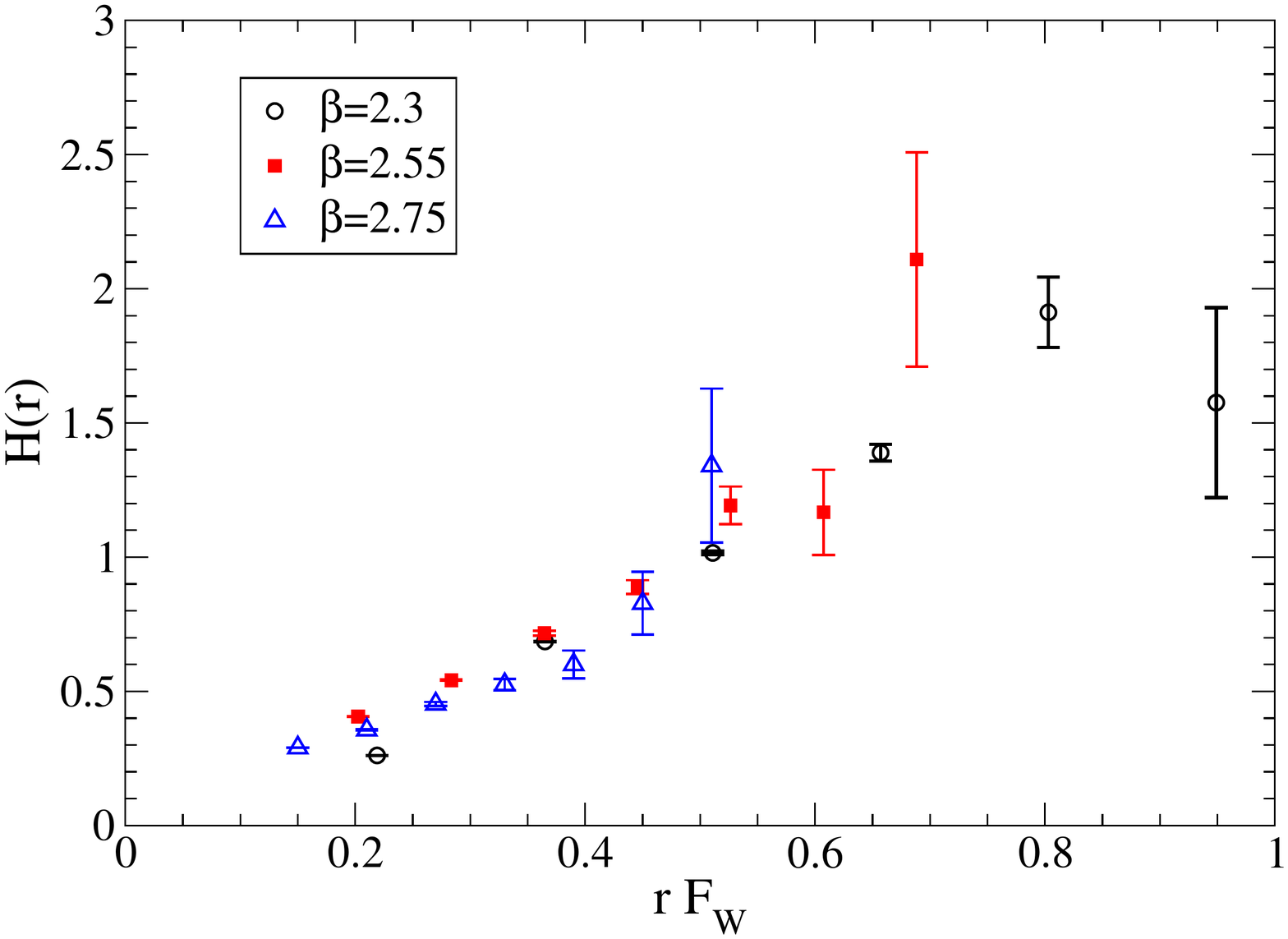} 
\caption{$H(r)$ as a function of $r$ in physical units of $F_W$ for the three $\beta$ values. }
\label{fig:h}
\end{center}
\end{figure}

We note that it is necessary to subtract the disconnected contribution, which does not vanish in this case, because the mass term breaks central charge conjugation. 
The result for $\beta=2.55$ is shown in the left panel of Figure~\ref{fig:potential}. There is a clear indication that the potential does not 
rise linearly after some distance. This is expected from the fact that static charges can be screened by the $\Omega$ fields, however the interesting result is that the scale at which this happens 
seems to also show scaling. This can be seen by plotting the function $H(r) \equiv r^2 {\partial V(r) \over \partial r}$ as a function of $r$ in physical units, see Figure~\ref{fig:h}. Within the 
large error bars, the curves seem to fall on a universal line for the three $\beta$. 

Since we have a mixing problem of stringy and static-light states the overlap of the Polyakov loop with the ground state is actually expected to depend on $r$.  
In order to simultaneously extract values for the ground  and first excited levels it would be necessary to consider other correlators such as those of smeared Wilson loops.  Asymptotically far from the mixing region, the ratio of two Polyakov loop correlators computed for different time extensions provides a more reliable determination of the static-light  
energy, as the dependence on the overlap of the Polyakov loop with the ground state drops out.  
Such ratio is shown in the right plot of Figure~\ref{fig:potential}, where the energy seems to reach a plateau at large $r$. 

 In a future study we plan to tune $\kappa$ to fix the ratio $m_W/F_W$, that is the effective coupling. In this case the ratio of the scalar to vector mass will be a prediction. 
 It will also be interesting to measure the self-couplings of the vector and those 
 of the vector to the scalar, as well as to search for higher resonances.  For this, as well as for having a more precise measurement
 of the static potential algorithmic improvements are necessary along the lines of \cite{Luscher:2001up}.

 A recent numerical study of this model at significantly smaller physical volumes can be found in \cite{Ferrari:2013lja}.

\section{Conclusions}
\label{sec:conclu}

We have studied a lattice formulation of the $SU(2)$ gauged non-linear chiral model, which is exactly equivalent to the Wilson formulation of an $SU(2)$ gauge theory with a mass term. The lattice theory has an exact $SU(2)$ global symmetry, and its associated conserved current has the quantum numbers of a vector triplet. Although the theory is not perturbatively renormalizable, we have considered the possibility that a scaling region might exist. Under this assumption, if the lightest state is a  vector triplet we have argued that the Wilsonian EFT describing the low-energy dynamics of this scaling region must be a massive gauge theory, even if confinement is at work . Other light states are expected therefore to unitarize the theory, such as a light scalar. 
We have carried out a numerical simulation of this model searching for lines of constant physics (as defined from the ratio of scalar to vector mass to have some fixed value) and we have found evidence for a scaling region, where however the effective coupling seems to be rather large, if the ratio $m_H/m_W$ is set to roughly its experimental value.
  These results suggests that this theory is a non-trivial strongly 
coupled theory that needs to be understood non-perturbatively.  It obviously has a clear physics motivation since it might be the simplest model for dynamical symmetry breaking. 

\section*{Acknowledgements}

We thank Claudio Pica for providing us a parallel $SU(2)$ gauge update code, from which we started writing our own routines.
MDM wishes to thank Jambul Gegelia for illuminating discussions. We also would like to thank Andrea Donini for a critical reading of the manuscript. 
This work was partially supported by the Spanish Minister of Education and Science projects RyC-2011-08557, FPA2011-29678, the Generalitat Valenciana (PROMETEO/2009/116), the Consolider-Ingenio CUP (CSD2008-00037) and CPAN(CSC2007-00042) and the European project ITN INVISIBLES (Marie Curie Actions, PITN-GA-2011-289442). 
The simulations have been carried out in the TIRANT cluster at the University of Valencia, member of the Spanish supercomputing network (RES).

\newpage

\appendix 

\section*{Appendix A. Peturbation theory at one-loop}

Although the lattice model is not perturbatively renormalizable, due to the $\Omega^\dagger \Omega =1$ constraint, a perturbative expansion is possible for small external momenta if $p^2 \ll  {\kappa\over a^2}$, or in other words for $\kappa$ large compared to any other scale in lattice units, and for small $g_0$.  

Let us discuss the range of validity of the expansion. There are four scales in this problem. The presence of the $\kappa$ term 
gives at tree level a mass to the lattice gauge fields:
\begin{eqnarray}
(m a)^2 = {\kappa g^2_0\over 2}.  
\end{eqnarray}
The equivalent of the vev of the Higgs is simply $ \kappa/a^2 = {2 m^2 \over g^2_0}$. Both scales being physical should be much 
smaller than the cutoff, therefore we should have $m^2 a^2 \ll 1, v^2 a^2 = \kappa \ll 1$. The perturbative expansion on the other hand assumes
$g_0$ small and also that the external momenta $p^2 \ll  {2 m^2\over g^2_0}= \kappa/a^2$. Note however that  we can be in two different regimes that are in principle
perturbative: 1) $p^2 \ll m^2 \ll {2 m^2 \over g^2_0}$ or 2) $m^2 \ll p^2 \ll {2 m^2 \over g^2_0}$. In the second case, we can neglect 
$m^2$ in front of external momenta, therefore the computation of the one-loop correction to $g^2_0$ coming from the diagrams
involving the plaquette interactions only can be found in the literature.  We will restrict our attention to this case. 
In order to ease notation we will set $a=1$ throughout. 

 We have used the background field method introduced in \cite{Luscher:1995vs} for the lattice regularization.  Besides the quantum fields, we introduce two background fields $B_\mu(x)$ and $\omega(x)$ in the following way
\begin{eqnarray}
U_\mu(x) = e^{i g_0 A_\mu(x)} ~e^{ i B_\mu(x)}, \;\;\; \Omega(x)= e^{ i \phi(x)} ~e^{i \omega(x)}.
\end{eqnarray}
 The gauge fixing procedure 
follows closely the lattice version of $R_\xi$-gauge introduced in \cite{Luscher:1995vs} modified as it is done in the case of spontaneously broken gauge theories in the continuum. The gauge-fixing  term in the action reads:
\begin{eqnarray}
S_{gf}[B, A, \phi] = {1\over  \xi} \left( D^L_\mu A_\mu + {1 \over 2} \xi \kappa g_0 \phi \right)^2, 
\end{eqnarray} 
where $D^L_\mu A_\mu \equiv A_\mu(x) - e^{-i B_\mu(x- \hat{\mu})} A_\mu(x-\hat{\mu}) e^{i B_\mu(x- \hat{\mu})}$, and the term in $\phi$ is introduced so that the mixed terms in $A_\mu \phi$ cancel. Note that it does not depend on the background field $\omega$. 

The ghost action can be derived straightforwardly and does not depend on the $\omega$ field. As a result there are no new ghost contributions at one-loop with
respect to those already present in the pure gauge theory. The are also no contributions from the measure terms at this order. 

The required one-loop corrections to $g_0$ and $\kappa$, $\Delta g^2$ and $\Delta \kappa$ respectively, can be read from two-point functions of the background fields computed at one-loop:

\vspace{1cm}
\includegraphics[width=12cm,angle=0]{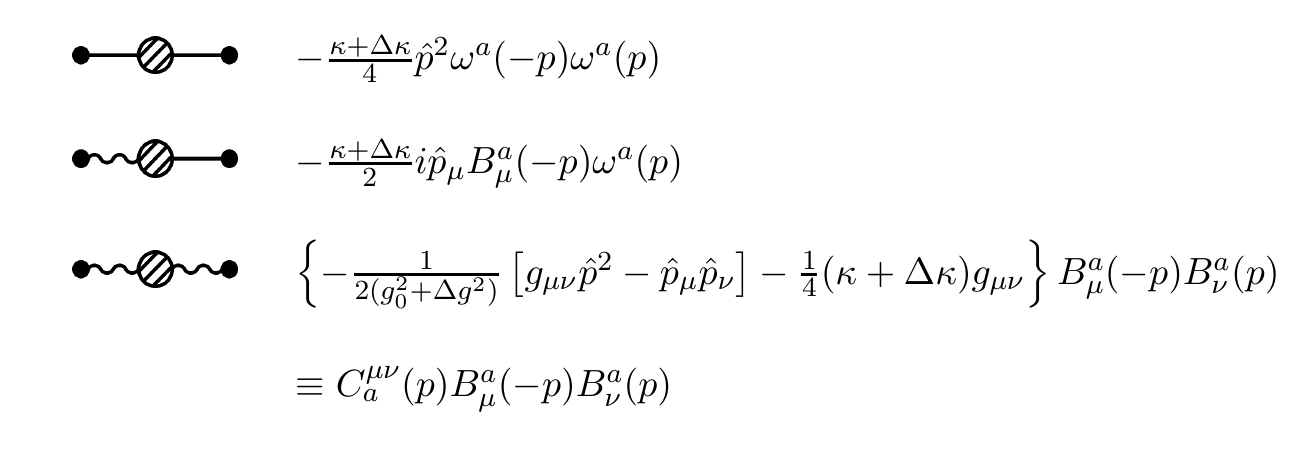} 

\begin{figure}
\begin{center}
\includegraphics[width=14cm,angle=0]{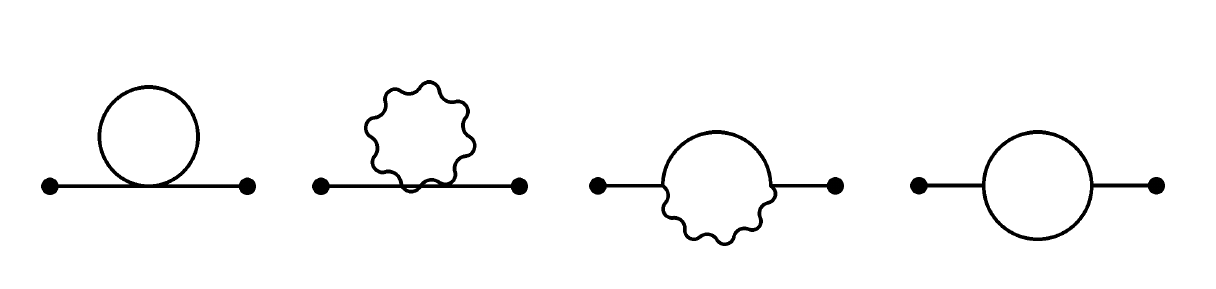} 
\caption{Diagrams contributing to $\omega$ two-point vertex function at one-loop. Solid, wiggly and dashed lines inside the loops correspond to $\phi$, $A_\mu$ and ghost fields respectively.}
\label{fig:ww}
\end{center}
\end{figure}

 The diagrams contributing to the two-point function of the $\omega$ field at one-loop are shown in Figure \ref{fig:ww}.

The correction to $\Delta \kappa$ can be
read directly from the coefficient of the one-loop $B \omega$ and $\omega \omega$ two-point vertices, but  also from the $p$-independent
term in  $BB$. The three results are the same. In Feynman gauge we obtain:
\begin{eqnarray}
&&{\Delta \kappa \over \kappa} = {1 \over \kappa} \left[ {1 \over 8 N}-{N\over 16} - {N\over 2} P_1 \right]\\
&&-g_0^2\left[\left({5 N\over 32} - {3 \over 16 N }\right) P_1 +{3 N \over 4} P_2 -{N\over 4} {1 \over (4 \pi)^2}- {3 N \over 4 (4 \pi)^2} \log(m^2)+{ 2 N \over (4 \pi)^2} F\left({m^2 \over p^2}\right) \right], \nonumber
\end{eqnarray}
where $m^2 \equiv {1 \over 2} \kappa g_0^2$, with the lattice integrals \cite{Luscher:1995np}
\begin{eqnarray}
P_1 &\equiv & \int_{-\pi}^{\pi} {d^4 p\over (2\pi)^4} {1 \over \hat{p}^2 }=   0.15493339...\nonumber\\
 P_2 &\equiv & \lim_{\mu \rightarrow 0} \left\{ {1 \over (4 \pi)^2}  \log(\mu^2) + \int_{-\pi}^\pi {d^4 p \over (2 \pi)^4} {1 \over (\hat{p}^2+ \mu^2)^2}\right\} = 0.02401318....
\end{eqnarray}
 The function $F(x)$ is given by
\begin{eqnarray}
F(x) \equiv 1 - \sqrt{1+ 4 x} ~\mathrm{arccoth} \sqrt{1+4 x}.
\end{eqnarray}

\begin{figure}
\begin{center}
\includegraphics[width=14cm,angle=0]{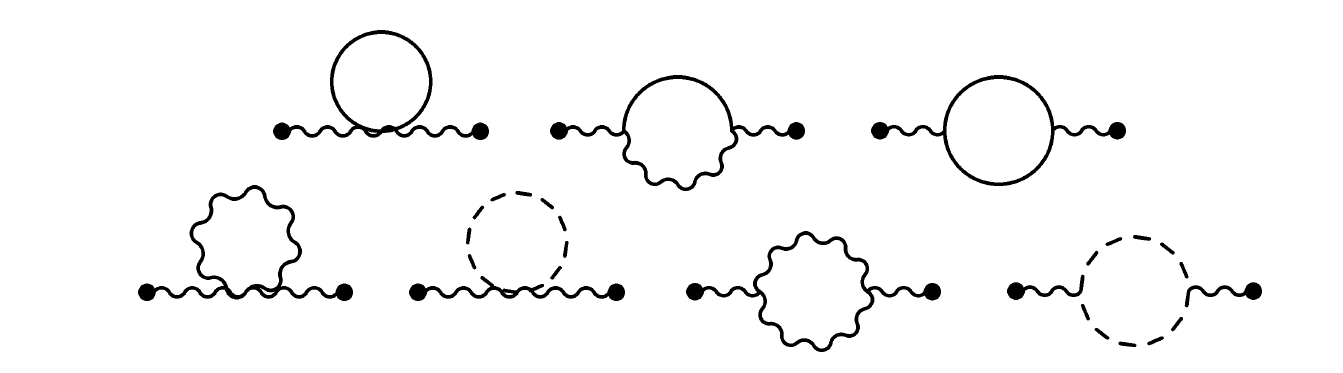} 
\caption{Diagrams contributing to the $B^a_\mu$ two-point function at one-loop. Solid, wiggly and dashed lines inside the loops correspond to $\phi$, $A_\mu$ and ghost fields respectively.}
\label{fig:bb}
\end{center}
\end{figure}

The one-loop correction to the gauge coupling gets new contributions proportional to $\kappa$,  corresponding to the diagrams in the first row of Figure~\ref{fig:bb}. For $\kappa=0$ only the diagrams in the second row of Figure~\ref{fig:bb} contribute. They have been computed in 
  \cite{Luscher:1995np} for massless gauge propagators. In our case, these diagrams are different because the gauge propagators 
  are massive. However, in the regime 
 $m^2 \ll p^2$ the results are the same up to corrections of ${\mathcal O}(m^2/p^2)$. 

 The contribution from the diagrams in  Figure~\ref{fig:bb} (up to ${\mathcal O}(m^2/p^2, a^2 p^2)$ corrections) is found to be:
 \begin{eqnarray}
  \left({\Delta g^2\over g_0^4}\right) = {N \over (4 \pi)^2} \left(-{29 \over 8} \ln p^2 + {63 \over 9} \right) + N  \left( {7\over 48} P_1 +{29\over 8} P_2 +{1\over 16} \right)-{1\over 8 N}.
 \end{eqnarray} 
 
 The $\beta$ function at one-loop is therefore
 \begin{eqnarray}
 \beta(g) & \equiv & - {\partial g \over \partial \ln a} = -{29 N \over 8 (4 \pi)^2 } g_0^3 +...,
 \end{eqnarray}
 There is asymptotic freedom. The coefficient coincides with that found in \cite{Fabbrichesi:2010xy} using a different regularization. 
 
  On the  other hand, if we want $m^2  \sim \kappa/a^2$ to be finite in the continuum limit, we need 
 that $\kappa + \Delta \kappa = {\mathcal O}(a^2)$, therefore  at the order we are working
 \begin{eqnarray}
 \kappa + \Delta \kappa = 0, 
 \end{eqnarray} 
 which is satisfied at the critical point 
 \begin{eqnarray}
 \kappa_c = { N \over 2} P_1 + {N\over 16} - {1 \over 8 N} + {\mathcal O}(g_0^2), 
 \end{eqnarray}
 which for $N=2$ corresponds to $\kappa_c= 0.2174+ {\mathcal O}(g_0^2)$. 
 Obviously $\kappa$ will get finite corrections to all orders and therefore the tuning must be done non-perturbatively.
 
 As it is well known not all the divergences can be reabsorbed in $g_0$ and $\kappa$. Some log's need to be absorbed in new couplings of ${\mathcal O}(p^4)$ operators \cite{Appelquist:1980ae}. 
 Although these divergences are logarithmic at one-loop, quadratic corrections to such coefficients will appear at higher orders of the loop expansion. It would seem 
  that a non-perturbative tuning is also required for these couplings to reach a scaling region. 
Whether at all, and in case at which values of the cutoff scale, this is necessary non-perturbatively is an open question and one of the motivations for the study presented
here. A perturbative analysis of this question in the context of a different regularization can be found in \cite{Bettinelli:2007cy}.


\begin{thebibliography}{10}

\bibitem{Aad:2012tfa}
{\bf ATLAS} Collaboration, G.~Aad et~al., {\it {Observation of a new particle
  in the search for the Standard Model Higgs boson with the ATLAS detector at
  the LHC}},  {\em Phys.Lett.} {\bf B716} (2012) 1--29,
  [\href{http://xxx.lanl.gov/abs/1207.7214}{{\tt arXiv:1207.7214}}].

\bibitem{Chatrchyan:2012ufa}
{\bf CMS} Collaboration, S.~Chatrchyan et~al., {\it {Observation of a new boson
  at a mass of 125 GeV with the CMS experiment at the LHC}},  {\em Phys.Lett.}
  {\bf B716} (2012) 30--61, [\href{http://xxx.lanl.gov/abs/1207.7235}{{\tt
  arXiv:1207.7235}}].

\bibitem{Englert:1964et}
F.~Englert and R.~Brout, {\it {Broken Symmetry and the Mass of Gauge Vector
  Mesons}},  {\em Phys.Rev.Lett.} {\bf 13} (1964) 321--323.

\bibitem{Guralnik:1964eu}
G.~Guralnik, C.~Hagen, and T.~Kibble, {\it {Global Conservation Laws and
  Massless Particles}},  {\em Phys.Rev.Lett.} {\bf 13} (1964) 585--587.

\bibitem{Higgs:1964pj}
P.~W. Higgs, {\it {Broken Symmetries and the Masses of Gauge Bosons}},  {\em
  Phys.Rev.Lett.} {\bf 13} (1964) 508--509.

\bibitem{PhysRevLett.19.1264}
S.~Weinberg, {\it A model of leptons},  {\em Phys. Rev. Lett.} {\bf 19} (Nov,
  1967) 1264--1266.

\bibitem{Wilson:1974sk}
K.~G. Wilson, {\it {Confinement of Quarks}},  {\em Phys.Rev.} {\bf D10} (1974)
  2445--2459.

\bibitem{Elitzur:1975im}
S.~Elitzur, {\it {Impossibility of Spontaneously Breaking Local Symmetries}},
  {\em Phys.Rev.} {\bf D12} (1975) 3978--3982.

\bibitem{Frohlich:1981yi}
J.~Fr{\"{o}}hlich, G.~Morchio, and F.~Strocchi, {\it {Higgs phenomenon without
  symmetry breaking order parameter}},  {\em Nucl.Phys.} {\bf B190} (1981)
  553--582.

\bibitem{Frohlich:1980gj}
J.~Fr{\"{o}}hlich, G.~Morchio, and F.~Strocchi, {\it {Higgs phenomenon without
  a symmetry breaking order parameter}},  {\em Phys.Lett.} {\bf B97} (1980)
  249.

\bibitem{Caudy:2007sf}
W.~Caudy and J.~Greensite, {\it {On the ambiguity of spontaneously broken gauge
  symmetry}},  {\em Phys.Rev.} {\bf D78} (2008) 025018,
  [\href{http://xxx.lanl.gov/abs/0712.0999}{{\tt arXiv:0712.0999}}].


\bibitem{Fradkin:1978dv}
E.~H. Fradkin and S.~H. Shenker, {\it {Phase Diagrams of Lattice Gauge Theories
  with Higgs Fields}},  {\em Phys.Rev.} {\bf D19} (1979) 3682--3697.

\bibitem{Lang:1981qg}
C.~Lang, C.~Rebbi, and M.~Virasoro, {\it {The phase structure of a nonabelian
  gauge Higgs field system}},  {\em Phys.Lett.} {\bf B104} (1981) 294.

\bibitem{Gegelia:2012yq}
J.~Gegelia, {\it {Why gauge symmetry ?}},
  \href{http://xxx.lanl.gov/abs/1207.0156}{{\tt arXiv:1207.0156}}.

\bibitem{Osterwalder:1973dx}
K.~Osterwalder and R.~Schrader, {\it {Axioms for Euclidean Green's functions}},
   {\em Commun.Math.Phys.} {\bf 31} (1973) 83--112.

\bibitem{Osterwalder:1974tc}
K.~Osterwalder and R.~Schrader, {\it {Axioms for Euclidean Green's Functions.
  2.}},  {\em Commun.Math.Phys.} {\bf 42} (1975) 281.

\bibitem{Luscher:1988uq}
M.~L{\"{u}}scher and P.~Weisz, {\it {Scaling Laws and Triviality Bounds in the
  Lattice phi**4 Theory. 3. N Component Model}},  {\em Nucl.Phys.} {\bf B318}
  (1989) 705.

\bibitem{Osterwalder:1977pc}
K.~Osterwalder and E.~Seiler, {\it {Gauge Field Theories on the Lattice}},
  {\em Annals Phys.} {\bf 110} (1978) 440.

\bibitem{Langguth:1985dr}
W.~Langguth, I.~Montvay, and P.~Weisz, {\it {Monte Carlo study of the standard
  SU(2) Higgs model}},  {\em Nucl.Phys.} {\bf B277} (1986) 11.

\bibitem{Campos:1997dc}
I.~Campos, {\it {On the SU(2) Higgs phase transition}},  {\em Nucl.Phys.} {\bf
  B514} (1998) 336--354, [\href{http://xxx.lanl.gov/abs/hep-lat/9706020}{{\tt
  hep-lat/9706020}}].


\bibitem{Bonati:2009pf}
C.~Bonati, G.~Cossu, M.~D'Elia, and A.~Di~Giacomo, {\it {Phase diagram of the
  lattice SU(2) Higgs model}},  {\em Nucl.Phys.} {\bf B828} (2010) 390--403,
  [\href{http://xxx.lanl.gov/abs/0911.1721}{{\tt arXiv:0911.1721}}].

\bibitem{Ferrari:2011aa}
R.~Ferrari, {\it {On the Phase Diagram of Massive Yang-Mills}},  {\em Acta
  Phys.Polon.} {\bf B43} (2012) 1965--1980,
  [\href{http://xxx.lanl.gov/abs/1112.2982}{{\tt arXiv:1112.2982}}].

\bibitem{Bettinelli:2012qk}
D.~Bettinelli and R.~Ferrari, {\it {On the Weak Coupling Limit for Massive
  Yang-Mills}},  {\em Acta Phys.Polon.} {\bf B44} (2013) 177--193,
  [\href{http://xxx.lanl.gov/abs/1209.4834}{{\tt arXiv:1209.4834}}].

\bibitem{Forster:1980dg}
D.~Forster, H.~B. Nielsen, and M.~Ninomiya, {\it {Dynamical Stability of Local
  Gauge Symmetry: Creation of Light from Chaos}},  {\em Phys.Lett.} {\bf B94}
  (1980) 135.

\bibitem{Fabbrichesi:2010xy}
M.~Fabbrichesi, R.~Percacci, A.~Tonero, and O.~Zanusso, {\it {Asymptotic safety
  and the gauged SU(N) nonlinear $\sigma$-model}},  {\em Phys.Rev.} {\bf D83}
  (2011) 025016, [\href{http://xxx.lanl.gov/abs/1010.0912}{{\tt
  arXiv:1010.0912}}].

\bibitem{D'Adda:1978uc}
A.~D'Adda, M.~L{\"{u}}scher, and P.~Di~Vecchia, {\it {A 1/n Expandable Series
  of Nonlinear Sigma Models with Instantons}},  {\em Nucl.Phys.} {\bf B146}
  (1978) 63--76.

\bibitem{D'Adda:1978kp}
A.~D'Adda, P.~Di~Vecchia, and M.~L{\"{u}}scher, {\it {Confinement and Chiral
  Symmetry Breaking in CP**n-1 Models with Quarks}},  {\em Nucl.Phys.} {\bf
  B152} (1979) 125--144.

\bibitem{Balachandran:1978pk}
A.~Balachandran, A.~Stern, and C.~Trahern, {\it {Nonlinear models as gauge
  thories}},  {\em Phys.Rev.} {\bf D19} (1979) 2416.

\bibitem{Bando:1984ej}
M.~Bando, T.~Kugo, S.~Uehara, K.~Yamawaki, and T.~Yanagida, {\it {Is rho Meson
  a Dynamical Gauge Boson of Hidden Local Symmetry?}},  {\em Phys.Rev.Lett.}
  {\bf 54} (1985) 1215.

\bibitem{Georgi:1989gp}
H.~Georgi, {\it {New realization of chiral symmetry}},  {\em Phys.Rev.Lett.}
  {\bf 63} (1989) 1917--1919.

\bibitem{Casalbuoni:1986vq}
R.~Casalbuoni, S.~De~Curtis, D.~Dominici, and R.~Gatto, {\it {Physical
  Implications of Possible J=1 Bound States from Strong Higgs}},  {\em
  Nucl.Phys.} {\bf B282} (1987) 235.

\bibitem{luscher:1998pe}
M.~L{\"{u}}scher, {\it {Advanced lattice QCD}},
  \href{http://xxx.lanl.gov/abs/hep-lat/9802029}{{\tt hep-lat/9802029}}.

\bibitem{Albanese:1987ds}
{\bf APE} Collaboration, M.~Albanese et~al., {\it {Glueball Masses and String
  Tension in Lattice QCD}},  {\em Phys.Lett.} {\bf B192} (1987) 163--169.

\bibitem{DellaMorte:2008jd}
M.~Della~Morte and L.~Giusti, {\it {Symmetries and exponential error reduction
  in Yang-Mills theories on the lattice}},  {\em Comput.Phys.Commun.} {\bf 180}
  (2009) 819--826, [\href{http://xxx.lanl.gov/abs/0806.2601}{{\tt
  arXiv:0806.2601}}].

\bibitem{DellaMorte:2010yp}
M.~Della~Morte and L.~Giusti, {\it {A novel approach for computing glueball
  masses and matrix elements in Yang-Mills theories on the lattice}},  {\em
  JHEP} {\bf 1105} (2011) 056, [\href{http://xxx.lanl.gov/abs/1012.2562}{{\tt
  arXiv:1012.2562}}].

\bibitem{Wolff:2003sm}
{\bf ALPHA} Collaboration, U.~Wolff, {\it {Monte Carlo errors with less
  errors}},  {\em Comput.Phys.Commun.} {\bf 156} (2004) 143--153,
  [\href{http://xxx.lanl.gov/abs/hep-lat/0306017}{{\tt hep-lat/0306017}}].

\bibitem{Luscher:2001up}
M.~L{\"{u}}scher and P.~Weisz, {\it {Locality and exponential error reduction
  in numerical lattice gauge theory}},  {\em JHEP} {\bf 0109} (2001) 010,
  [\href{http://xxx.lanl.gov/abs/hep-lat/0108014}{{\tt hep-lat/0108014}}].

\bibitem{Ferrari:2013lja}
  R.~Ferrari,
    Acta Phys.\ Polon.\ B {\bf 44} (2013) 9,  1871--1885,
  [arXiv:1308.1111].


\bibitem{Luscher:1995vs}
M.~L{\"{u}}scher and P.~Weisz, {\it {Background field technique and
  renormalization in lattice gauge theory}},  {\em Nucl.Phys.} {\bf B452}
  (1995) 213--233, [\href{http://xxx.lanl.gov/abs/hep-lat/9504006}{{\tt
  hep-lat/9504006}}].

\bibitem{Luscher:1995np}
M.~L{\"{u}}scher and P.~Weisz, {\it {Computation of the relation between the
  bare lattice coupling and the MS coupling in SU(N) gauge theories to two
  loops}},  {\em Nucl.Phys.} {\bf B452} (1995) 234--260,
  [\href{http://xxx.lanl.gov/abs/hep-lat/9505011}{{\tt hep-lat/9505011}}].

\bibitem{Appelquist:1980ae}
T.~Appelquist and C.~W. Bernard, {\it {The nonlinear sigma model in the loop
  expansion}},  {\em Phys.Rev.} {\bf D23} (1981) 425.


\bibitem{Bettinelli:2007cy}
  D.~Bettinelli, R.~Ferrari and A.~Quadri,
  Phys.\ Rev.\ D {\bf 77} (2008) 045021
  [arXiv:0705.2339] and Phys.\ Rev.\ D {\bf 77} (2008) 105012
   [Erratum-ibid.\ D {\bf 85} (2012) 129901]
  [arXiv:0709.0644]. 
  
  
    
\end{thebibliography}

\providecommand{\href}[2]{#2}\begingroup\raggedright\endgroup

\end{document}